\documentclass[review,authoryear,3p,times,10pt]{elsarticle}

\usepackage{multirow,setspace,times,amssymb,amsmath,graphicx,color,rotating,subfigure,url}
\usepackage{lineno,color}
\usepackage{natbib}
\usepackage{booktabs}%
\usepackage{longtable}%
\graphicspath{{Figures/}}
\usepackage[table]{xcolor}
\usepackage{tabularx}
\usepackage{graphicx} 
\usepackage{epstopdf}
\usepackage{mathrsfs}
\usepackage{makecell}
\usepackage{threeparttable}
\usepackage[bookmarks=true,colorlinks,linkcolor=blue,anchorcolor=blue,citecolor=blue,unicode]{hyperref}
\usepackage{bookmark}
\usepackage{booktabs}
\usepackage[T1]{fontenc}
\usepackage{array,caption,threeparttable}
\usepackage[font=small,labelfont=bf,labelsep=none]{caption}
\hypersetup{CJKbookmarks=true}%
\journal{Finance Research Letters}
\begin{document}

\begin{frontmatter}
\title{Spillover effects between climate policy uncertainty, energy markets, and food markets: A time-frequency analysis}

\author[HNUST]{Ting Zhang}
\ead{tzhang@hunst.edu.cn}
\author[HNUST]{Peng-Fei Li}
\author[SB,RCE,DM]{Wei-Xing Zhou\corref{CorAuth}}
\ead{wxzhou@ecust.edu.cn}
\cortext[CorAuth]{Corresponding author. Corresponding to: 130 Meilong Road, P.O. Box 114, School of Business, East China University of Science and Technology, Shanghai 200237, China.}

\address[HNUST]{School of Business, Hunan University of Science and Technology, Xiangtan 411201, China}
\address[SB]{School of Business, East China University of Science and Technology, Shanghai 200237, China}
\address[RCE]{Research Center for Econophysics, East China University of Science and Technology, Shanghai 200237, China}
\address[DM]{Department of Mathematics, East China University of Science and Technology, Shanghai 200237, China}

\begin{abstract}
The study examines the return connectedness between climate policy uncertainty (CPU), clean energy, fossil energy, and food markets. Using the time-domain method of \cite{Diebold-Yilmaz-2012-IntJForecast} and frequency-domain methods of \cite{Barunik-Kehlik-2018-JFinancEconom}, we find  substantial spillover effects between these markets. Furthermore, high frequency domain is the primary driver of overall connectedness. In addition, CPU is a net contributor of return shocks in the short term, whereas it turns to be a net recipient in the medium and long terms. Across all frequencies, clean energy and oils are consistent net recipients, while meat is a dominant net contributor. 

\end{abstract}


\end{frontmatter}


\section{Introduction}

The global climate crisis, including shifting weather patterns, rising sea levels, and extreme weather events, poses significant risks to economies and societies worldwide. To address these challenges, policymakers have introduced several initiatives such as the Paris Agreement and the European Green Deal, alongside emissions reduction targets set by COP26 \citep{Khalfaoui-MeftehWali-Viviani-BenJabeur-Abedin-Lucey-2022-TechnolForecastSocChang,Karim-Naeem-Shafiullah-Lucey-Ashraf-2023-FinancResLett}. However, the implementation of these measures has introduced substantial climate policy uncertainty (CPU), which affects various markets, particularly energy and food markets \citep{Guo-Long-Luo-2022-IntRevFinancAnal}. The formulation and implementation of climate policies have curtailed the consumption of fossil energy to some degree while accelerating the growth of the clean energy industry \citep{Su-Yuan-Tao-Shao-2022-TechnolForecastSocChang}. In addition, climate and environmental changes impact food markets since a certain level of carbon dioxide growth can raise grain yield and then subsequently affects food prices \citep{Han-Hua-Engel-Guan-Yin-Wu-Sun-Wang-2022-AgricWaterManage}.

The relationship between energy and food prices has long been a focus of researchers and policymakers \citep{Youssef-Mokni-2021-IntEconJ}. Researchers identify three main channels linking energy and food markets. First, changes in production costs driven by energy prices directly affect food prices \citep{Georgiou-Acha-Shah-Markides-2018-JCleanProd}. Second, rising biofuel demand caused by the high fossil fuel prices will then raise the prices of the raw materials like corn and soybeans \citep{Yoon-2022-RenewEnergy,Tanaka-Guo-Wang-2023-GCBBioenergy}. Third, energy and food prices are affected by the financial factors since the financialization of energy and food commodities. Also, as two categories of financial assets, food and energy prices are connected because of the hedging behaviors of cross-market investors \citep{Adil-Bhatti-Waqar-Amin-2022-JPublicAff}. While research on the food-energy nexus is growing, most studies primarily focus on the relationship between food and oil prices \citep{Mohammed-2022-OPECEnergyRev,Yu-Peng-Zakaria-Mahmood-Khalid-2023-JBusEconManag}, with fewer addressing other fossil energy like coal,natural gas, and gasoline \citep{Diab-Karaki-2023-EnergyEcon,Vatsa-Miljkovic-Baek-2023-JAgricEcon,Miljkovic-Vatsa-2023-IntRevFinancAnal}. However, the role of clean energy remains underexplored, despite climate change and CPU accelerate the transformation of energy applied in agricultural production from fossil energy to clean energy.

Existing research demonstrates that CPU significantly impacts energy markets by influencing both consumption and prices. For instance, \cite{Syed-Apergis-Goh-2023-Energy} and \cite{Addey-Nganje-2024-EnergyEcon} find that CPU inhibits clean energy consumption, while \cite{Shang-Han-Gozgor-Mahalik-Sahoo-2022-RenewEnergy} and \cite{Zhou-Siddik-Guo-Li-2023-RenewEnergy} suggest that CPU have positive effects on clean energy consumption in specific contexts. Moreover, CPU affects energy prices and the performance of energy stocks and bonds, with previous studies providing nonlinear and time-varying evidences \citep{Iqbal-Bouri-Shahzad-Alsagr-2024-EnergyEcon,Bouri-Iqbal-Klein-2022-FinancResLett}. In food market, CPU influence agricultural production and food prices \citep{Laborde-Mamun-Martin-Pineiro-Vos-2021-NatCommun,Guo-Li-Zhang-Ji-Zhao-2023-JCommodMark}. Researchers find significant spillover effects between CPU and key commodity prices such as soybeans and corn \citep{,Wang-Wang-Yunis-Kchouri-2023-EnergyEcon}. 

Although a large number of studies have provided valuable insight into the relationships between CPU, energy, and food markets, several research gaps remain. From the perspective of research objects, most studies on the food-energy nexus focus on fossil energy market, and there is limited studies on clean energy. Furthermore, while the effects of CPU on food and energy markets have been studied independently, little attention has been given to their interconnectedness as a system \citep{Yan-Cheung-2023-FinancResLett, Gong-Jin-Liu-2023-Energy}. From the methodology point of view, previous research primarily uses time-domain analyses methodologies, such as TVP-VAR-DY model \citep{Vo-Tran-2024-TechnolForecastSocChang,AhmadianYazdi-Roudari-Omidi-Mensi-AlYahyaee-2024-IntRevEconFinanc,Tanaka-Guo-Wang-2024-EnergyEcon}, with limited research explores the frequency-domain connectedness. In this study, we contribute to the existing literature by incorporating both fossil energy and clean energy markets, along with subcategories of food market, including meat, dairy, cereals, oils, and sugar, to explore their heterogeneous associations. Furthermore, we examine spillovers between CPU, food, and energy markets from both time- and frequency-domain perspectives, providing valuable insights for policymakers, investors, and researchers.

The remainder of this paper is organized as follows. Section~\ref{S2:Methodology:Data} describes the methodologies and data. Section~\ref{S3:EmpAnal} presents the empirical results and discussions, and Section~\ref{S4:Conclusion} concludes.

\section{Methodology and Data}
\label{S2:Methodology:Data}

\subsection{Time-domain spillover index: TVP-VAR-DY model}

A notable empirical approach to examine the spillover effects among markets is the one by \cite{Diebold-Yilmaz-2012-IntJForecast}, who introduce the connectedness measure based on the generalized forecast error variance decomposition (GFEVD). Furthermore, \cite{Antonakakis-Chatziantoniou-Gabauer-2020-JRiskFinancManag} extend the \cite{Diebold-Yilmaz-2014-JEconom} by using a time-varying parameter vector autoregressive model (TVP-VAR) with a time-varying covariance structure.

Following \cite{Antonakakis-Chatziantoniou-Gabauer-2020-JRiskFinancManag}, TVP-VAR-DY model is shown as:
\begin{equation}
    \label{EQ:TVP1}
    Y_t=\sum_{i=1}^r\beta_tY_{t-1}+\epsilon_t, ~~\epsilon_t \sim N(0,S_t)
\end{equation}
\begin{equation}
    \label{EQ:TVP2}
    \beta_t=\beta_{t-1}+v_t,~~v_t\sim N(0,R_t)
\end{equation}
with $Y= \left[y_1, y_2, \cdots, y_M\right]^{\rm{T}}$ and $\beta^{\prime}_t=  \left[\beta_{1t},   \beta_{2t}, \cdots, \beta_{rt} \right]^{\rm{T}}$,
where $Y_t$ is an $M\times1$ vectors and $y_1, y_1,\cdots,y_M$ represent the individual markets. $r$ is the optimal lag length of TVP-VAR-DY model obtained by AIC. $\epsilon_t$ is a $M \times 1$ vector, $v_t$ is an $M^2r \times 1$ vector, and they are the noise terms. $S_t$ and $R_t$ are the time-varying variance-covariance matrices. The generalized forecast error variance decompositions in $H$-step ahead as follows:
\begin{equation}
    \label{EQ:TVP3}
    d_{ij}(h) = \frac{\sigma^{-1}_{jj}\sum^{H-1}_0(e^{\prime}_iA_h\sum e_j)^2}{\sum^{H-1}_0(e^{\prime}_iA_h\sum A{\prime}_he_i)}
\end{equation}
where $\sum$ is the covariance matrix of $\epsilon_t$, and $\sigma^{-1}_{jj}$ is the standard error of $\epsilon_t$, $e_j$ is a vector with the $j$-th unit value 1, others value 0. 

The spillovers from market $i$ to market $j$ can be represented by the variance decomposition matrix $D_{ij}(h)$:
\begin{equation}
    \label{EQ:TVP4}
    D_{ij}(h) =
    \left \{ 
    \begin{matrix}
    d_{11}& d_{12} &\dots& d_{1M}\\
    d_{21}& d_{22} &\dots& d_{2M}\\
    \dots &\dots   &\dots& \dots\\
    d_{M1}& d_{M2} &\dots& d_{MM}
    \end{matrix}
    \right \}
\end{equation}

Furthermore, we standardized $d_{ij}(h)$ to $l_{ij}(h)$ to make sure that the sum of each row is 1:
\begin{equation}
    \label{EQ:TVP5}
    l_{ij}(h) = \frac{d_{ij}(h)}{\sum^M_{j=1}d_{ij}(h)}
\end{equation}

Then the total spillover index (TSI) is as follows:
\begin{equation}
    \label{EQ:TCI}
    TSI(h) = 100 \times \frac{\sum^M_{i,j=1,i\neq j}l_{ij}(h)}{\sum^M_{i,j=1}d_{ij}(h)}
\end{equation}

The directional spillover index that market $i$ transmit to all other markets and receive  from all other markets are shown as below:
 \begin{equation}
     \label{EQ:TO}
     TO_{i} = 100 \times \frac{\sum^M_{j=1,i\neq j}d_{ij}(h)}{\sum^M_{j=1}d_{ij}(h)}
 \end{equation}
  \begin{equation}
     \label{EQ:FROM}
     FROM_{i} = 100 \times \frac{\sum^M_{j=1,i\neq j}d_{ij}(h)}{\sum^M_{i,j=1}d_{ij}(h)}
 \end{equation}
 
The net spillover of market $i$ can be measured by the disparity between $TO_{i}$ and $FROM_{i}$:
 \begin{equation}
     \label{EQ:NET}
     NET_{i} = TO_{i} - FROM_{i}
 \end{equation}

We can further calculate the net pairwise directional spillover (NPDS) as follows:
 \begin{equation}
     \label{EQ:NPDC}
     NPDC_{i \leftarrow j} = l_{ij} - l_{ji}
 \end{equation}

\subsection{Frequency-domain spillover effects: TVP-VAR-BK model}

 Based on the \cite{Diebold-Yilmaz-2012-IntJForecast}'s spillover index, \cite{Barunik-Kehlik-2018-JFinancEconom} propose a spectral representation of the GFEVD results and calculate the connectedness between markets over different frequencies. 

For a frequency band $d = (a,b): a,b \in (-\pi, \pi), a < b$, the GFEVD on $d$ is expressed as:
\begin{equation}
    \label{EQ:BK1}
    (l_d)_{i,j} = \frac{1}{2\pi}\int_d l_{i,j}(\omega)d\omega
\end{equation}
where, 
\begin{equation}
    \label{EQ:BK1.1}
    l_{i,j}(\omega) = \frac{\theta_{i,j}(\omega)}{\sum^M_{j=1}\theta_{i,j}(\omega)}
\end{equation}
where $\theta_{i,j}(\omega)$ is the portion of the spectrum of the $i$th variable at frequency $\omega$ that can be attributed to the impact of the $j$th variable.

Then we can compute the corresponding index of $TCI$, $TO$, $FROM$, $NET$, and $NPDC$ on frequency $d$.

\subsection{Data}

The Climate Policy Uncertainty (CPU) index used in this paper is constructed by \cite{Gavriilidis-2021-SSRN}\footnote{Data of CPU index is obtained from \url{https://www.policyuncertainty.com}}. We employ the Food Price Index (FPI) released by FAO\footnote{The website address of FAO is \url{https://www.fao.org/}} to track the price changes of the food markets, including meat, dairy, cereals, oils, and sugar. The subcategories are selected according to FAO's index data release, and they are also essential for global food security. Referring to \cite{Chen-Liang-Ding-Liu-2022-EnergyEcon} and \cite{Saeed-Bouri-Alsulami-2021-EnergyEcon}, we use the iShares US Oil \& Gas Exploration \& Production ETF (IEO) as the representative measure of fossil energy market and the iShares Global Clean Energy ETF (ICLN) as the proxy of clean energy market\footnote{Data on these ETFs is extracted from the Wind Database (\url{https://www.wind.com.cn/}).}. The sample period spans from January 2012 to December 2023, and all data used in this study are at a monthly frequency. The time span is determined by data availability and encompasses a critical phase of global climate policy development.

\begin{table*}[!h]
  \centering
  \setlength{\abovecaptionskip}{0.1cm}
  \caption{Descriptive statistics.}
     \setlength{\tabcolsep}{0.9mm}{
    \begin{threeparttable}
    \begin{tabular}{cccccccccccc}
    \toprule
     & CPU & Fossil Energy & Clean Energy  & Meat & Dairy  & Cereals  & Oils & Sugar \\
     \midrule
Mean& 0.0076 & 0.0037 & -0.0005 & 0.0005 & -0.0002 & -0.0002 & -0.0012 & -0.0006 \\
Median& -0.0160 & 0.0031 & 0.0033 & 0.0009& 0.0029 & -0.0033& -0.0081 & -0.0017 \\
Std.Dev.& 0.3753 & 0.1020 & 0.0898 & 0.0199 & 0.0338 & 0.0342 & 0.0554 & 0.0617 \\
Skewness& 0.1512 & -0.7542 & -1.7873 & -0.1303 & 0.0558 & 0.8085 & 0.1069 & -0.1777 \\
Kurtosis& 3.3763 & 10.1492 & 12.9431 & 2.8938 & 3.4845 & 7.4900 & 5.6264 & 4.1455 \\
Jarque-Bera& 1.3888$^{*}$ & 318.0900$^{***}$ & 665.2$^{***}$ & 0.47179  & 1.4729$^{*}$ & 135.7$^{***}$ & 41.372$^{***}$ & 8.5702$^{**}$  \\
ADF& -8.1284$^{***}$ & -4.7320$^{***}$ & -4.9140$^{***}$ & -6.2571$^{***}$ & -3.5698$^{**}$ & -4.6030$^{***}$ & -4.4390$^{***}$  & -4.2101$^{***}$  \\
PP& -21.616$^{***}$ & -13.324$^{***}$ & -12.958$^{***}$ & -6.2571$^{***}$  & -6.9328$^{***}$ & -9.0698$^{***}$ & -8.4755$^{***}$ & -9.7653$^{***}$  \\
KPSS& 0.0182 & 0.1091 & 0.1812 & 0.0654 & 0.0947 & 0.2112 & 0.1542 & 0.2570  \\
ZA&-18.1026$^{***}$ & -14.1091$^{***}$ & -13.6841$^{***}$ & -6.3985$^{***}$ & -7.4007$^{***}$ & -10.0135$^{***}$ & -9.7801$^{***}$ & -10.2973$^{***}$  \\
    \bottomrule
    \end{tabular}
    \begin{tablenotes}
    \footnotesize
    \item Notes: ADF, PP, ZA, and KPSS are the empirical statistics of the augmented Dickey-Fuller (1979) and Phillips and Perron (1988), ZA (Zivot and Andrew, 1992) unit root tests, and the Kwiatkowski et al. (1992) stationarity test, respectively. * denotes significant at 1 \% level of significance.
    \end{tablenotes}
    \end{threeparttable}
    }
  \label{tab:statistics}
\end{table*}

\begin{figure}[!h]
\centering
\includegraphics[width=0.9\linewidth]{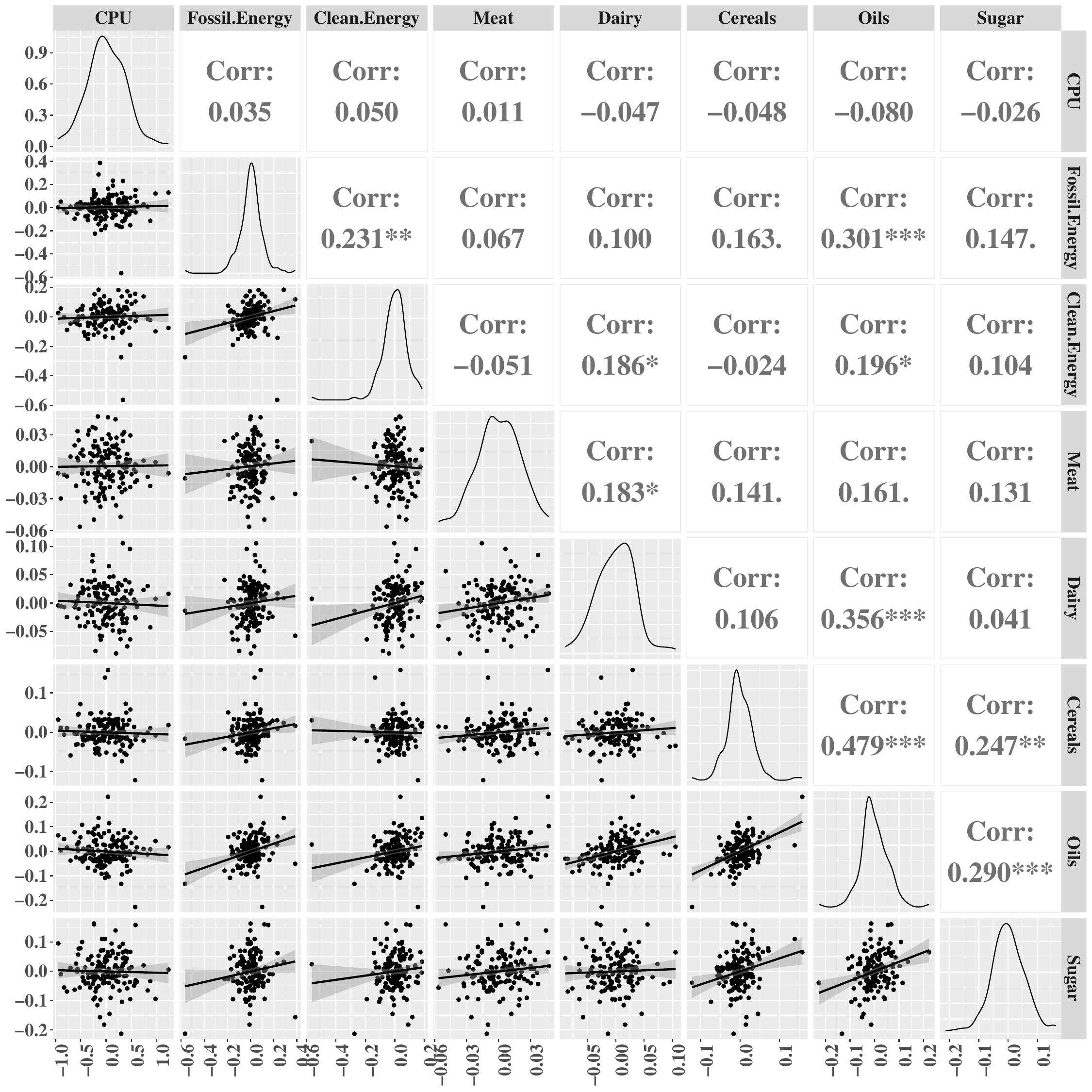}
\caption{Correlation analysis. Notes: Lower part of the figure is the scatterplots of each pair of variables. Upper part is the correlation coefficients. Distributions are presented on the diagonal.} 
\label{Fig:ScatterMatrix}
\end{figure}

Table~\ref{tab:statistics} reports the descriptive statistics and the unit root test results for the log returns of these indices. CPU exhibits the highest volatility, while food markets show lower volatility levels compared to CPU and energy markets. This suggests that food prices tend to be more stable than energy prices, likely due to government interventions in agricultural markets. The skewnesses are reported to be positive for CPU, dairy, cereals, and oils, while they are negative for fossil energy, clean energy, meat, and sugar. Meat is platykurtic, while the remaining variables are leptokurtic. All variables except meat are non-normally distributed according to the Jarque-Bera statistics, implying the presence of asymmetric and extreme price movements in these markets. They are all stationary as shown by various unit root tests like ADF, ZA, PP, and KPSS, indicating that it is valid to apply the TVP-VAR models. In addition, Fig.~\ref{Fig:ScatterMatrix} shows that most variables are positively correlated with the exception of CPU and food markets.

\section{Results and Discussions}
\label{S3:EmpAnal}

\subsection{Static spillover effects}
\label{S3.1:Static SE}

Table~\ref{tab:Static spillovers} presents the static spillover index calculated by the \cite{Diebold-Yilmaz-2012-IntJForecast} and the \cite{Barunik-Kehlik-2018-JFinancEconom} method. The connectedness is decomposed into three distinct frequency bands, i.e., one to four months (short term), four to twelve months (medium term), and twelve to infinite months (long term), and it is denoted as $TCI_d$ on the bands of $d_1 \in \left[\frac{\pi}{4}, \pi \right]$, $d_2 \in \left[\frac{\pi}{12}, \frac{\pi}{4} \right]$ and $d_3 \in \left[0, \frac{\pi}{12} \right]$.

From the DY result in Table~\ref{tab:Static spillovers}, the total spillover index between CPU, energy, and food markets is 25.28\%. According to the ranking of 'To' and 'From', oil market emerges as the largest contributor to spillover effects, followed by fossil energy and cereal markets. They are also the top three recipients. CPU ranks the lowest in the system, both as a contributor and a recipient. Moreover, fossil energy and clean energy markets are net contributors, whereas CPU is a net recipient. Within the food market, cereal and oil markets are net contributors, while meat, dairy, and sugar markets are net recipients. The substantial spillover effects from cereals and oils to other markets can be attributed to their dual roles as both food commodities and biofuel inputs, which makes them highly sensitive to energy price fluctuations and policy shifts \citep{Myers-Johnson-Helmar-Baumes-2014-AmJAgrEcon,Yoon-2022-RenewEnergy}.

From the frequency-domain perspective, the total spillover index is 14.67\% in the short term, decreasing to 6.13\% in the medium term and 4.48\% in the long term. Spillovers are predominated by the short-term domain and weaken as the frequency increases. This suggests that these markets process information rapidly, so that shocks to any market in the system are transmitted to other markets generally within four months \citep{Wang-2020-EnergyEcon}. In the short term, oil market is the largest contributor, and fossil energy market is the largest recipient. It supports previous studies of \cite{Wu-Ren-Wan-Liu-2023-FinancResLett} which find that fossil energy receives risks from agricultural markets. In the medium term, oil market remains the largest contributor and recipient. Similarly, in the long term, oil market is still the largest contributor, while meat and oil markets are the top recipients. These results underscore the critical role of oil market in the system, which is consistent with broadly recognized findings in the literature \citep{Menier-Bagnarosa-Gohin-2024-ApplEcon,Issariyakul-Dalai-2014-RenewSustEnergRev,Peri-Baldi-2010-EnergyEcon}.

\begin{table*}[!t]
  \centering
  \setlength{\abovecaptionskip}{0.1cm}
  \caption{Static spillovers: Overall and frequency domains}
    \begin{tabular}{cccccccccc}
    \toprule
    \textbf{DY} & CPU  & Fossil Energy  & Clean Energy & Meat & Dairy & Cereals & Oils & Sugar  & From\\
    CPU                 & 91.35	& 2.20 & 2.88 &	0.30 &	0.58 &	0.21& 0.94 & 1.54& 8.65 \\
    Fossil Energy  & 0.44&70.49 &11.50 &	1.20 &	2.78 &	1.36 &	7.70& 4.53& 29.51 \\
    Clean Energy  & 2.66& 11.04  & 73.66&0.18 & 4.56 &	0.14 & 5.07& 2.70& 26.34 \\
    Meat               & 1.73& 3.26   & 0.89  & 76.77 & 1.31 & 11.23 & 3.20  & 1.62  & 23.23 \\
    Dairy              & 0.39& 1.41   & 4.65  & 3.19   & 81.13 & 0.71  & 8.46  & 0.07  & 18.87 \\
    Cereals            & 0.92&1.02 & 0.76 &	2.57 & 2.12&	68.47 & 18.78& 5.37& 31.53 \\
    Oils               & 0.48&8.97 & 4.87&	1.78&	7.30&	12.55 & 60.75& 3.31& 39.25 \\
    Sugar              & 1.31&	6.28 &	3.56&	0.64&	0.07&	5.52&	7.47& 75.17& 24.83 \\
   To  & 7.92&	34.17&	29.09&	9.86&	18.72&	31.71&	51.61& 19.14& \multirow{2}{1cm}{TCI= \\ 25.28}\\
       \vspace{10bp}
    Net &-0.73&	4.66&	2.75 &	-13.37&	-0.15 &	0.18 & 12.36& -5.70 \\
    \textbf{BK-Short} & CPU  & Fossil Energy  & Clean Energy & Meat & Dairy & Cereals & Oils & Sugar  & From\\
    CPU                 & 82.52	& 2.19 & 2.61 &	0.08 &	0.39 &	0.18& 0.91 & 1.51& 7.87\\
    Fossil Energy  & 0.29&57.76 &10.15 &	0.41 &	1.29 &	0.55 &	5.40& 4.39& 22.48 \\
    Clean Energy  & 1.66& 9.01  & 53.96&0.09 & 1.78 &	0.08 & 3.37& 2.51& 18.50\\
    Meat               & 1.35& 1.62  & 0.28   & 27.16 & 0.88 & 1.90 & 0.90& 1.07& 7.99 \\
    Dairy              & 0.11 & 0.74  & 1.59   & 0.71   & 36.13& 0.12 & 3.28& 0.05& 6.61 \\
    Cereals            & 0.63&0.67 & 0.54 &	1.91 & 0.71&	44.58 & 9.38& 4.99& 18.83 \\
    Oils               & 0.34&4.47 & 2.57&	1.16&	2.88&	8.77 & 36.91& 2.16& 22.35 \\
    Sugar               & 0.80&2.92 &	1.86&	0.13&	0.03&	3.54&	3.47& 52.84& 12.74 \\
    To  & 5.19&	21.63&	19.60&	4.50&	7.95&	15.13&	26.71& 16.69 & \multirow{2}{1cm}{TCI= \\ 14.67}\\
        \vspace{10bp}
    Net &-2.69&	0.85&	1.10 &	-3.49&	1.34 &	-3.71 & 4.35& 3.94 \\
      \textbf{BK-Medium}  & CPU  & Fossil Energy  & Clean Energy & Meat & Dairy & Cereals & Oils & Sugar  & From\\
    CPU                 & 6.06	& 0.00 & 0.20 &	0.11 &	0.11 &	0.01& 0.01 & 0.03& 0.48 \\
    Fossil Energy  & 0.08&8.33 &0.80 &	0.41 &	0.81 &	0.40 &	1.33& 0.12& 3.97 \\
    Clean Energy  & 0.64& 1.32  & 12.79&0.05 & 1.59 &	0.05 & 1.04& 0.14& 4.82 \\
    Meat               & 0.26  & 0.93 & 0.29   & 26.02& 0.11 & 4.54 & 1.09 & 0.40& 7.63 \\
    Dairy              &0.15    & 0.38 & 1.67   & 1.17  & 25.20& 0.27 & 2.80 & 0.02& 6.45 \\
    Cereals            & 0.18&0.22 & 0.14 &	0.56 & 0.79&	15.89 & 5.85& 0.32& 8.07\\
    Oils               & 0.10&2.80 & 1.40&	0.34&	2.45&	2.39 & 15.03& 0.76& 10.23\\
    Sugar            & 0.33&	2.09 &	1.07&	0.24&	0.02&	1.16&	2.46& 14.71& 7.37\\
    To  & 1.74&	7.75&	5.57&	2.87&	5.89&	8.83&	14.59& 1.79 &  \multirow{2}{1cm}{TCI= \\ 6.13}\\
        \vspace{10bp}
    Net &1.25&	3.78&	0.75 &	-4.75&	0.57 &	0.76 & 4.36& -5.58 \\
    \textbf{BK-Long}  & CPU  & Fossil Energy  & Clean Energy & Meat & Dairy & Cereals & Oils & Sugar  & From\\
    CPU                  & 2.77& 0.00 & 0.07 &	0.11 &	0.08 &	0.01& 0.02 & 0.01& 0.29 \\
    Fossil Energy  & 0.06 &1.40 &0.54 &0.38 &	0.68 &	0.41 &	0.97& 0.01& 3.06 \\
    Clean Energy  & 0.36& 0.71  & 6.90&0.04 & 1.19 &	0.01 & 0.66& 0.05& 3.03\\
    Meat               & 0.11 & 0.72  & 0.31& 23.60 & 0.31 & 4.79   & 1.21& 0.15  & 7.61 \\
    Dairy              & 0.13  & 0.29  & 1.38& 1.30   & 19.80& 0.31  & 2.38& 0.01& 5.80 \\
    Cereals           & 0.11&	0.12 & 0.08 &	0.10 & 0.62&	8.01 & 3.54& 0.05& 4.63 \\
    Oils              & 0.04&	1.69 & 0.90&	0.29&	1.98&	1.40 & 8.81& 0.38& 6.67 \\
    Sugar             & 0.18&	1.27 &	0.63&	0.26&	0.02&	0.82&	1.53& 7.62& 4.72 \\
    To  & 0.99&	4.79	&3.92&	2.49&	4.88&	7.75&	10.32& 0.66  & \multirow{2}{1cm}{TCI= \\ 4.48}\\
    Net &0.70&	1.73&	0.89 &	-5.12&	-0.93 &	3.13 &  3.65& -4.06 \\

    \bottomrule
    \end{tabular}
  \label{tab:Static spillovers}
\end{table*}

\subsection{Network analysis}
\label{S3.2:Network}
\begin{figure}[!h]
   \centering
	\includegraphics[width=0.49\linewidth]{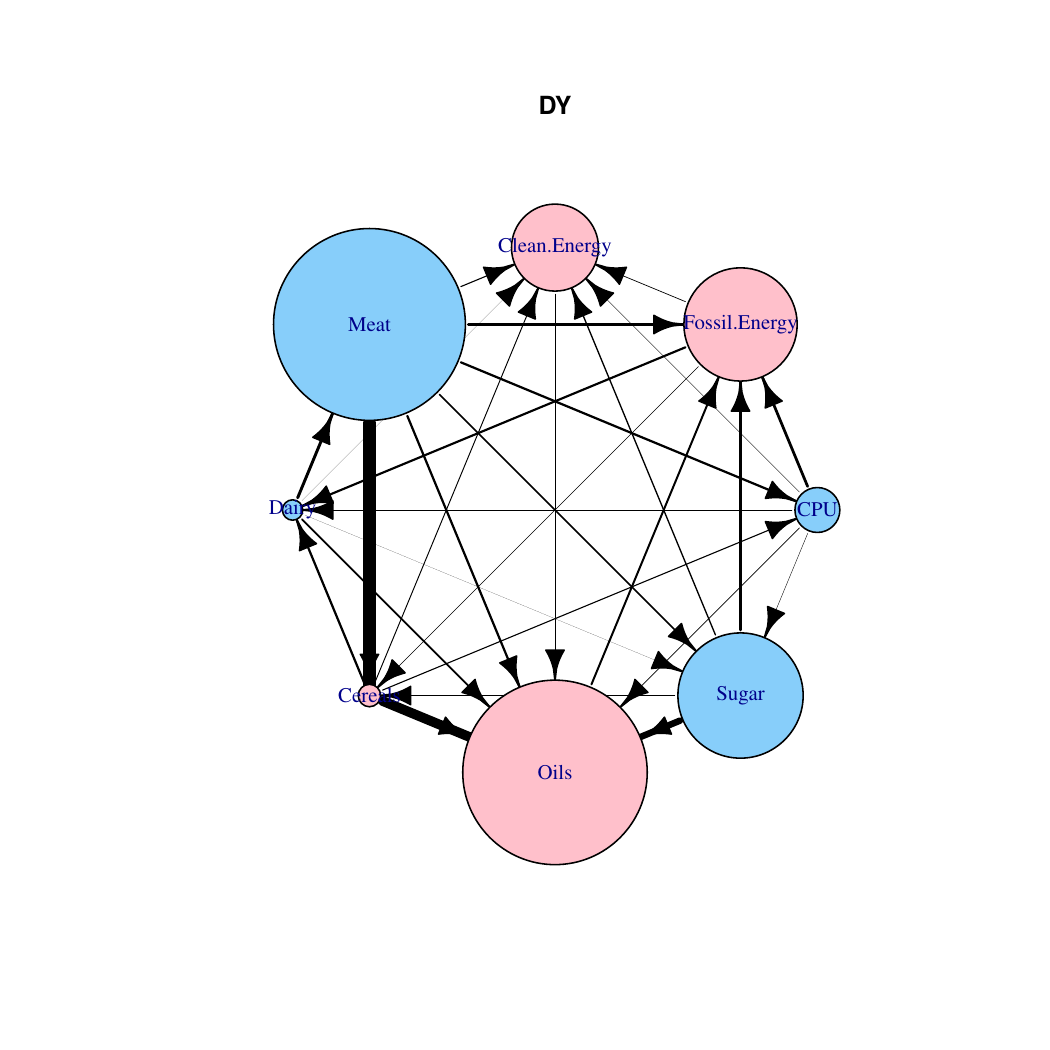}
	\includegraphics[width=0.49\linewidth]{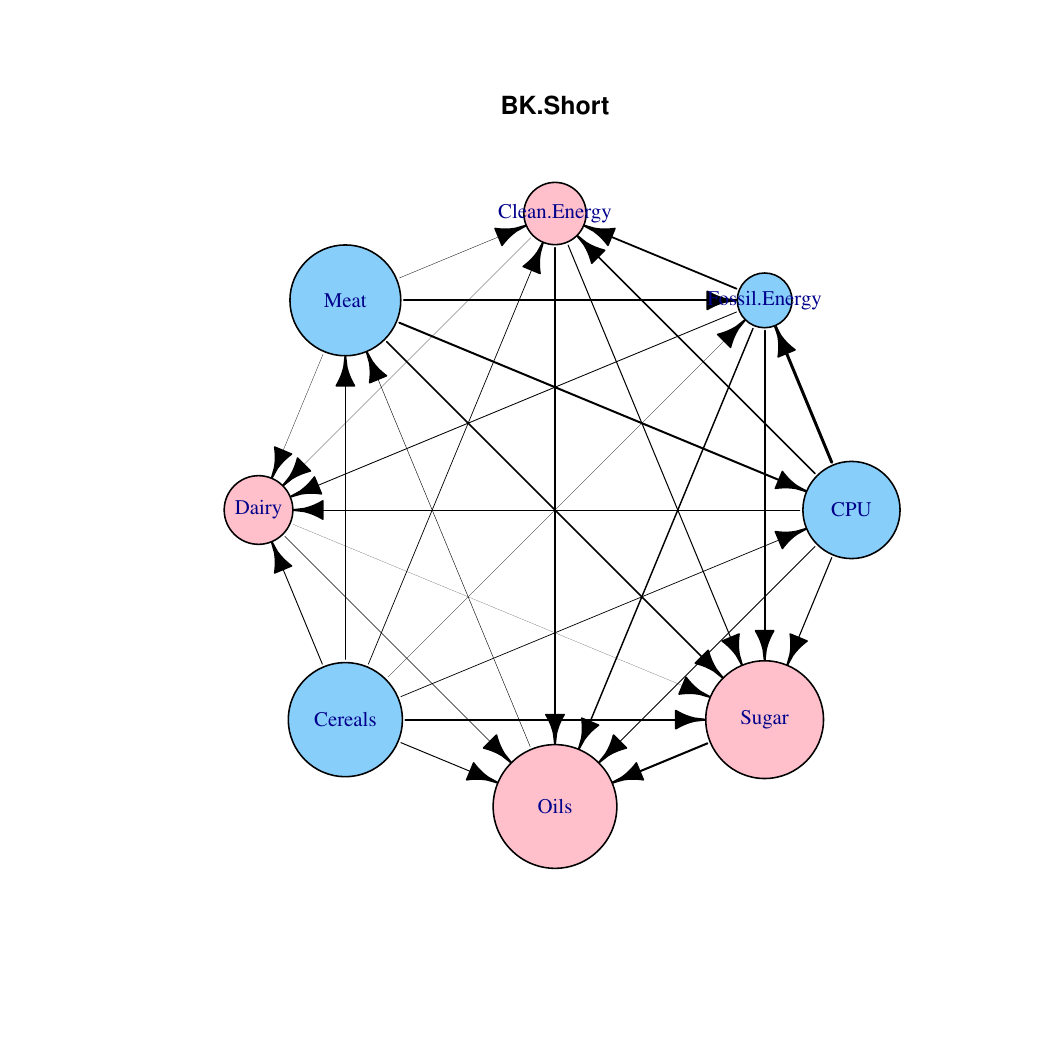}
	\includegraphics[width=0.49\linewidth]{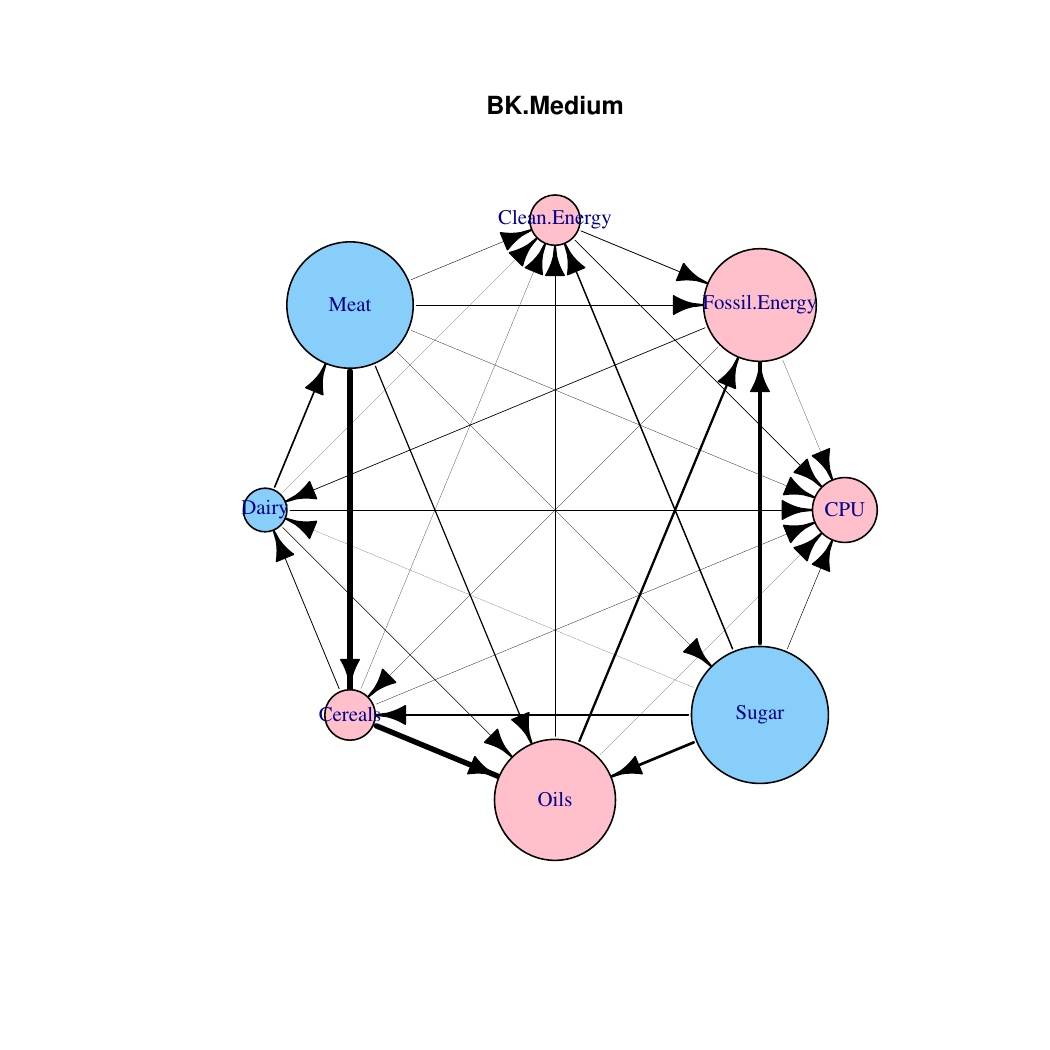}
	\includegraphics[width=0.49\linewidth]{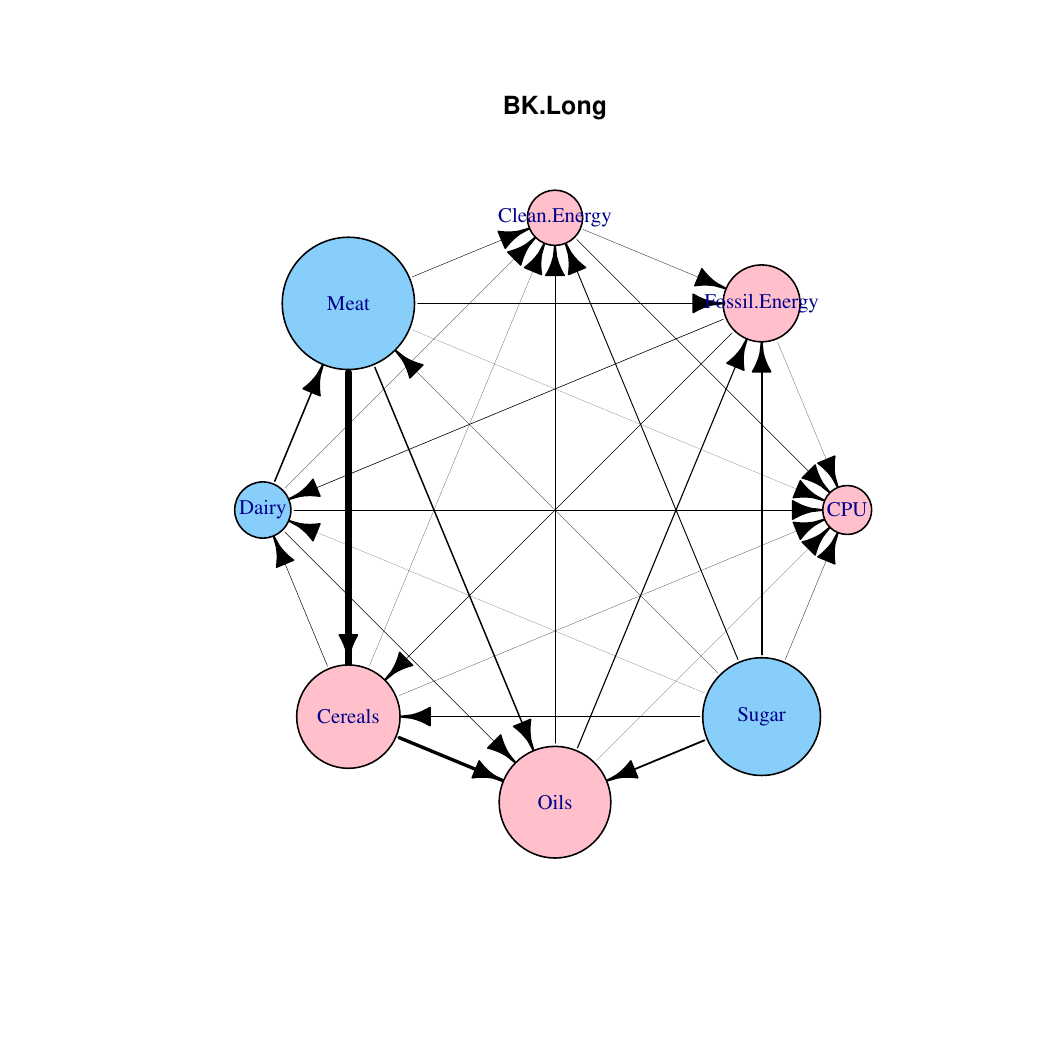}
	\vspace{-30bp}
    \caption{Network of pair-wise net spillovers. Notes: The size of the node represents the absolute value of the net spillover index. Pink indicates a net receiver, and blue indicates a net transmitter. The arrow and line thickness indicate the direction and size of the net pairwise spillover index, respectively. } 
    \label{Fig:Network}
\end{figure}

To further visualize the net spillover relationships between these markets, we present the network diagrams of net spillover for total and frequency bands in Fig.~\ref{Fig:Network}. In the DY results, clean energy is a net recipient of shocks from all markets except oils, while oil market only transmits shocks to fossil energy market. This pattern may be related to the development of biofuels \citep{TaghizadehHesary-Rasoulinezhad-Yoshino-2019-EnergyPolicy}. The dominant routes include transmissions from meat to cereals and from cereals to oils. CPU is a net transmitter in the short term and turns to a net receiver as the time horizon lengthens. Specifically, in the short term, CPU transmits risks to clean energy, fossil energy, dairy, oils, and sugar, while receiving risks from meat and cereals. However, in the medium- and long-term, CPU is a net receiver for all energy and food markets. The short-term transmissions of risks from CPU to other markets can be explained by the immediate effects of climate policies, such as regulatory changes, emissions restrictions, or the introduction of subsidies for clean energy. These policy actions have rapid impacts on energy and food markets by affecting investor expectations, technological advancements, or production costs \citep{Kettunen-Bunn-Blyth-2011-EnergyJ}. On the other hand, in the medium- and long-term, the role of CPU shifts to a net receiver due to the cumulative and delayed effects of climate policies. Over time, these policies lead to structural changes in the economy and then introduce new sources of uncertainty and market instability, which makes CPU more sensitive to spillovers from other markets \citep{Qiao-Chang-Mai-Dang-2024-EnergyEcon,Siddique-Nobanee-Hasan-Uddin-Hossain-Park-2023-EnergyEcon}.

Table~\ref{tab: Centrality rankings} presents the centrality rankings based on degree, closeness, betweenness, and eigenvector centrality measures at short-, medium-, and long-term. Across all measures, meat market consistently emerges as one of the most central nodes, particularly under degree and closeness centrality rankings. Oil market consistently emerges as a key node in eigenvector centrality across all horizons, reflecting its high interconnectedness with other critical nodes in the network.

\begin{table*}[!t]
  \centering
  \setlength{\abovecaptionskip}{0.1cm}
  \caption{Centrality rankings}
     \resizebox{\textwidth}{35mm}{
    \begin{tabular}{llccccccccc}
    \toprule
    & & 1& 2& 3 & 4 & 5& 6 & 7 & 8\\
     \midrule
    \multirow{4}{*}{\textbf{DY}} & \textbf{Degree} & Meat & Dairy & CPU & Fossil Energy & Sugar & Cereals & Clean Energy & Oils \\
    & \textbf{Closeness} & Clean Energy & Dairy & Sugar & Cereals & CPU & Oils & Fossil Energy & Meat \\
    & \textbf{Between} & Cereals & CPU & Dairy & Fossil Energy & Clean Energy & Oils & Sugar & Meat \\
    & \textbf{Eigenvector} & Oils& Fossil Energy & Cereals & Dairy & Clean Energy  & Meat & CPU & Sugar \\
     \midrule
    \multirow{4}{*}{\textbf{BK-Short}} & \textbf{Degree} & Meat & CPU &  Fossil Energy &  Clean Energy & Dairy & Sugar & Oils & Cereals \\
    &\textbf{Closeness} & Meat & Cereals & Dairy & Sugar & Clean Energy & Fossil Energy & Oils & CPU \\
    & \textbf{Between} & Meat & Oils & Dairy & CPU & Fossil Energy & Clean Energy & Cereals & Sugar \\
    & \textbf{Eigenvector} & Oils & Sugar & Clean Energy & Fossil Energy & Dairy & Meat & CPU & Cereals \\
     \midrule
    \multirow{4}{*}{\textbf{BK-Medium}} &\textbf{Degree} & Meat & Sugar & Cereals & Dairy & Oils & Fossil Energy & Clean Energy & CPU \\
    &\textbf{Closeness} & CPU & Dairy & Sugar & Clean Energy & Fossil Energy & Oils & Cereals & Meat \\
    & \textbf{Between} & Dairy & Fossil Energy & Clean Energy & Cereals & Meat & Sugar & CPU & Oils \\
    & \textbf{Eigenvector} & Oils & Fossil Energy & Cereals & Dairy & Clean Energy & CPU & Meat & Sugar \\
     \midrule
    \multirow{4}{*}{\textbf{BK-Long}}& \textbf{Degree} & Sugar & Meat &  Cereals & Dairy & Oils & Fossil Energy & Clean Energy & CPU \\ 
    &\textbf{Closeness} & CPU & Meat & Oils & Dairy & Sugar & Fossil Energy & Cereals & Clean Energy \\
    & \textbf{Between} & Fossil Energy & Dairy & Cereals & Clean Energy & CPU & Oils & Meat & Sugar \\
    & \textbf{Eigenvector} & Oils & Cereals & Fossil Energy & Dairy & Clean Energy & Meat & CPU & Sugar \\
    \bottomrule
    \end{tabular}
     }
  \label{tab: Centrality rankings}
\end{table*}

\subsection{Dynamic spillover effects}
\label{S3.2:Dynamic SE}

To investigate the time-varying spillover effect, we conduct the dynamic analysis by using the TVP-VAR model \citep{Antonakakis-Chatziantoniou-Gabauer-2020-JRiskFinancManag}.  Fig.~\ref{EQ:TCI} illustrates the dynamic total connectedness. As we see, the spillover effect between CPU, energy, and food markets is significant and fluctuates over time. The total spillover index reaches its peak at the end of 2012 and then keeps going down throughout the remainder of the sample period. This is in accordance with the finding of \cite{Kang-Tiwari-Albulescu-Yoon-2019-EnergyEcon} that the connectedness between crude oil and agriculture commodities peaks in 2012 because of the food crisis and the European debt crisis. In addition, the spillover effects are predominantly driven by high-frequency connectedness, with spillovers increasing with frequency.

\begin{figure}[!h]
   \centering
	\includegraphics[width=0.325\linewidth]{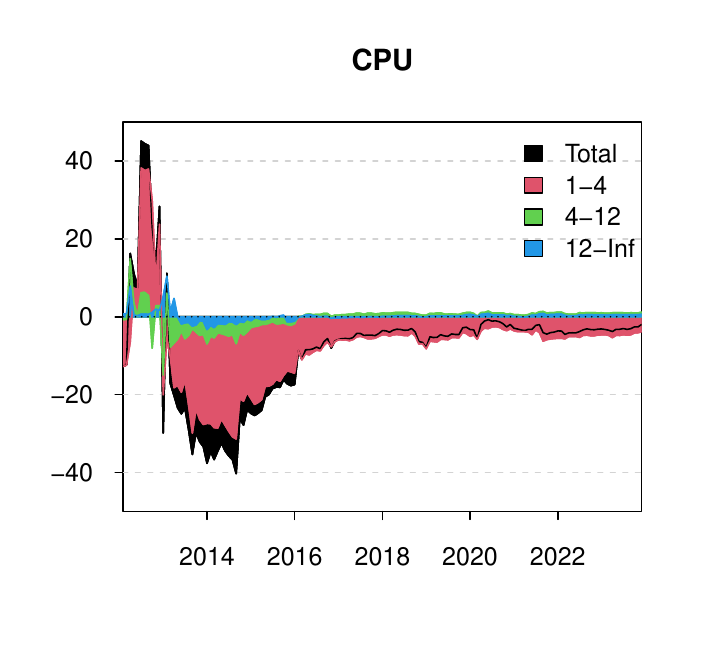}
	\includegraphics[width=0.325\linewidth]{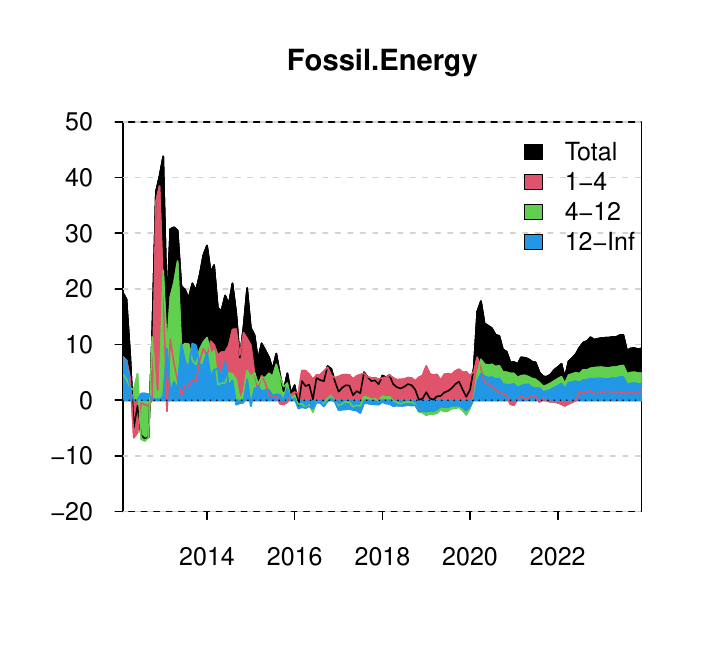}
	\includegraphics[width=0.325\linewidth]{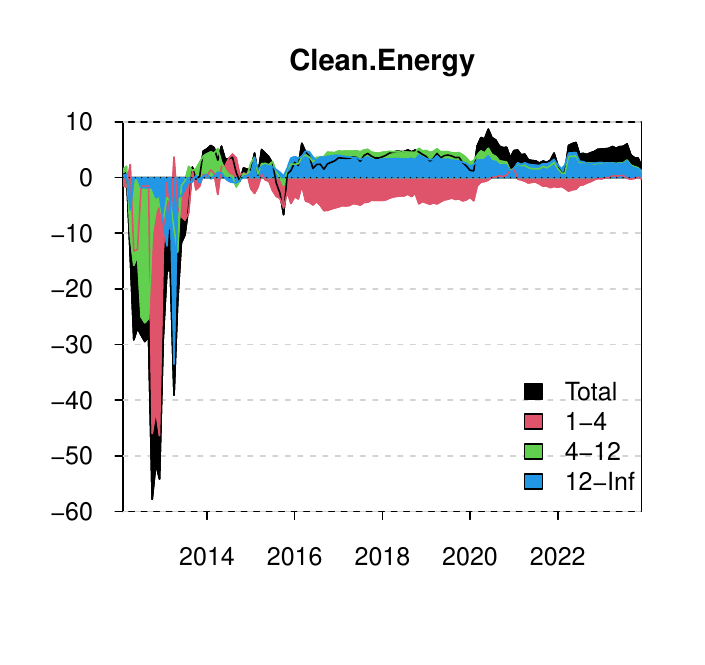}
	\includegraphics[width=0.325\linewidth]{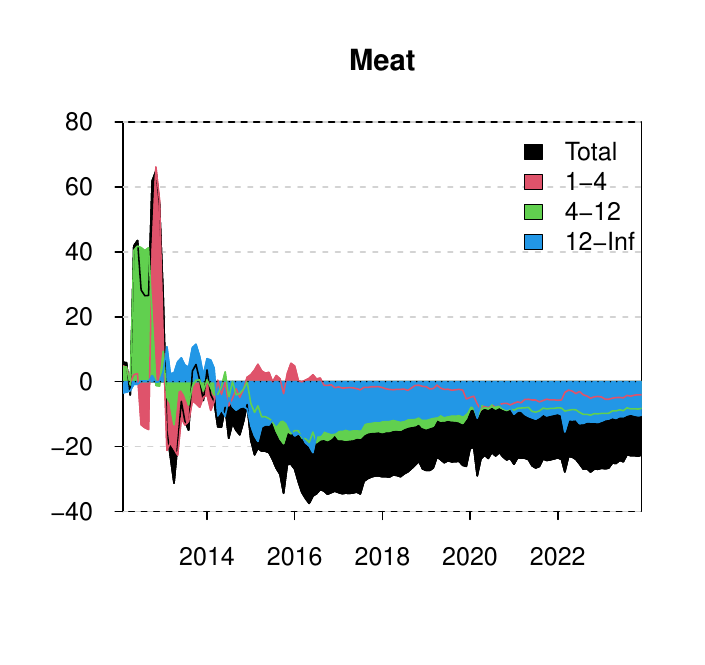}
	\includegraphics[width=0.325\linewidth]{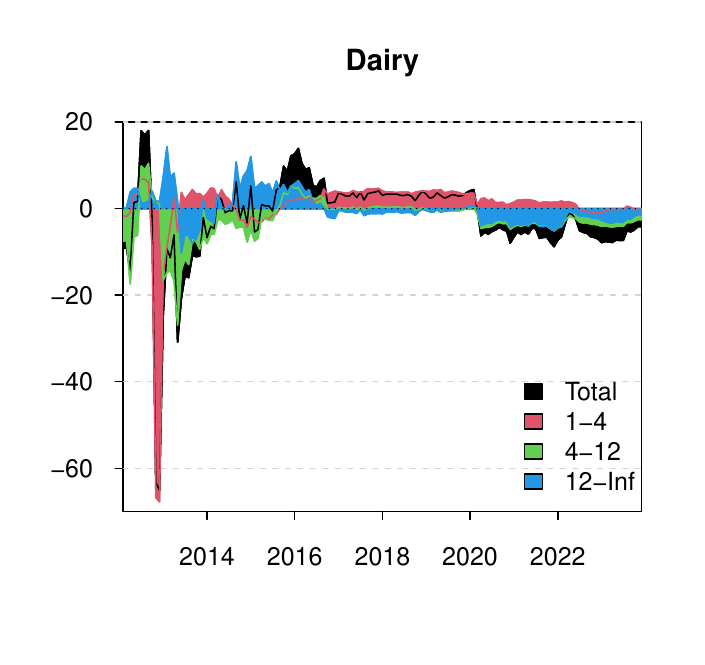}
	\includegraphics[width=0.325\linewidth]{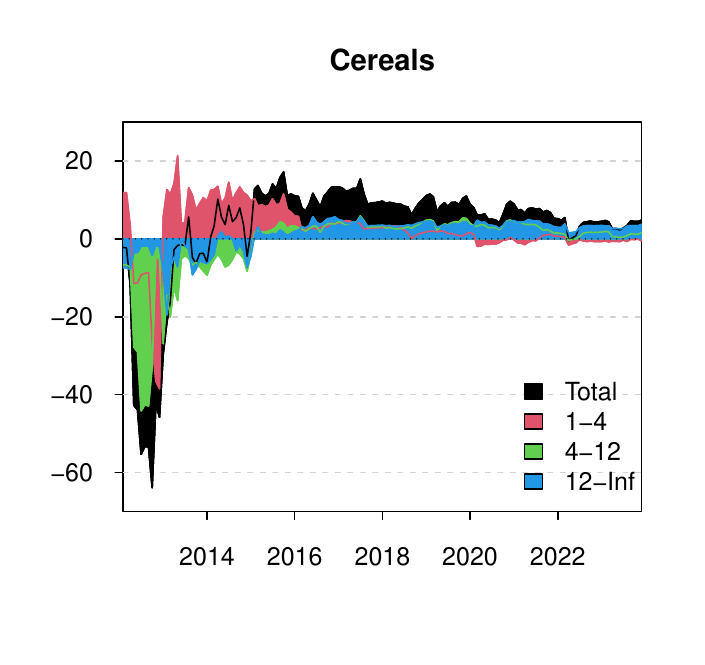}
	\includegraphics[width=0.325\linewidth]{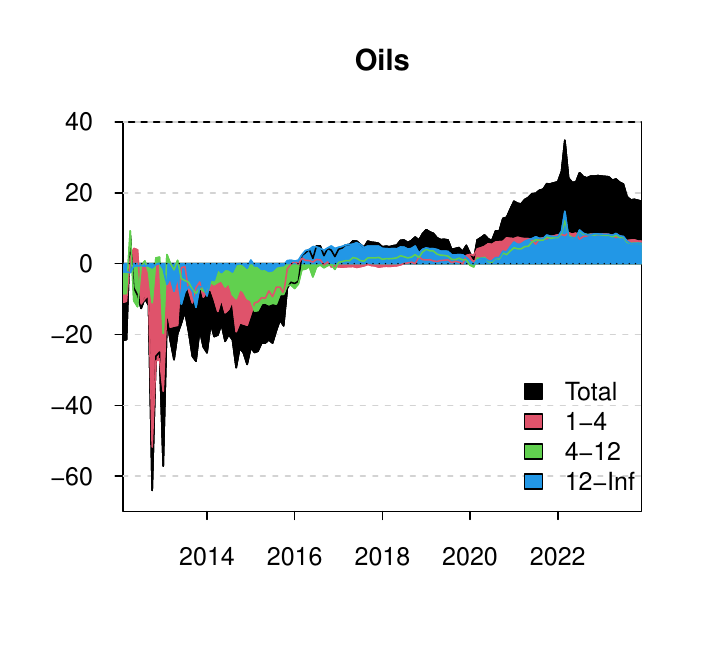}
	\includegraphics[width=0.325\linewidth]{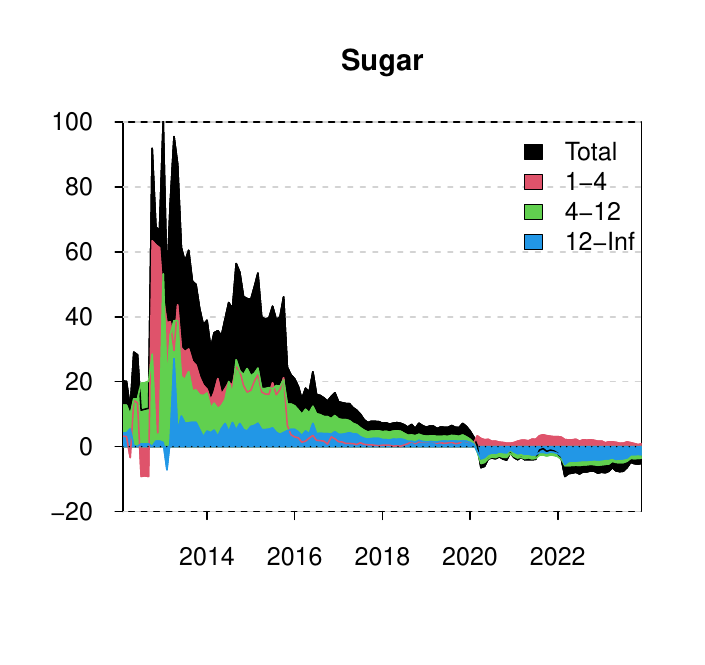}
	 \vspace{-20bp}
  \caption{The net spillover index.} 
    \label{Fig:NCI}
\end{figure}

Fig.~\ref{Fig:NCI} presents net spillovers. The net spillovers of all variables show a larger jump at the end of 2012. Subsequently, CPU plays the role of a net recipient, while energy markets are net contributors. For food markets, meat and cereal markets are net recipients, oil market consistently acts as a contributors. Notably, dairy and sugar markets shift from contributors to recipients in 2020.

\begin{figure}[!t]
\centering
\includegraphics[width=0.6\linewidth]{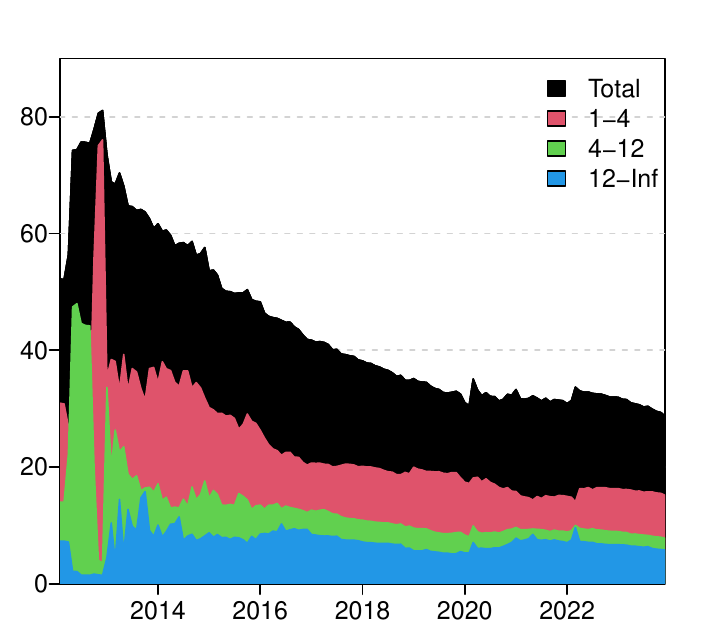}
\caption{The total spillover index.} 
\label{Fig:TCI}
\end{figure}

\section{Conclusion}
\label{S4:Conclusion}

This study investigates the return connectedness between CPU, energy, and food markets from the time and frequency perspectives by employing the spillover methodologies of \cite{Diebold-Yilmaz-2012-IntJForecast} and \cite{Barunik-Kehlik-2018-JFinancEconom}. The findings reveal significant spillover effects among these markets, with the high-frequency domain driving the majority of the overall connectedness. Moreover, CPU act as a net contributor in the short term, whereas it turns to a net recipient in the medium and long terms. Across all frequencies, clean energy and oil markets are net recipients, and meat market is a net contributor. Furthermore, oil market presents the highest centrality ranking in the spillover network, indicating the its importance in connecting the system. Dynamic analysis further highlights the persistent significance of spillover effects throughout the sample period, peaking at the end of 2012 and decline steadily after that. The evolution of the net directional spillover index illustrates that CPU and meat play more roles as net recipients, whereas energy markets, cereals, and sugar are net contributors for most of the sample period.

Our research has several practical implications. First, it reminds policymakers to prevent the short-term risk transmission from CPU to energy and food markets since CPU is a significant risks contributor at the high frequency band. Second, the return spillovers between CPU, energy, and food markets increase with the frequency bands, providing information on asset allocation and risk prevention for investors with different investment horizons. Third, the connectedness network helps cross-market investors and risk supervision departments identify the key nodes in risk transmission across markets. 

The CPU data used in this study is based on U.S. data, and future research can extend the analysis by incorporating CPU data from other countries, such as China and Australia, to examine their heterogeneous relationships with food and energy markets.



%
\bibliography{Bib2,Bib1}

\begin{thebibliography}{47}
\expandafter\ifx\csname natexlab\endcsname\relax\def\natexlab#1{#1}\fi
\providecommand{\url}[1]{\texttt{#1}}
\providecommand{\href}[2]{#2}
\providecommand{\path}[1]{#1}
\providecommand{\DOIprefix}{doi:}
\providecommand{\ArXivprefix}{arXiv:}
\providecommand{\URLprefix}{URL: }
\providecommand{\Pubmedprefix}{pmid:}
\providecommand{\doi}[1]{\href{http://dx.doi.org/#1}{\path{#1}}}
\providecommand{\Pubmed}[1]{\href{pmid:#1}{\path{#1}}}
\providecommand{\bibinfo}[2]{#2}
\ifx\xfnm\relax \def\xfnm[#1]{\unskip,\space#1}\fi
\bibitem[{Addey and Nganje(2024)}]{Addey-Nganje-2024-EnergyEcon}
\bibinfo{author}{Addey, K.A.}, \bibinfo{author}{Nganje, W.},
  \bibinfo{year}{2024}.
\newblock \bibinfo{title}{Climate policy volatility hinders renewable energy
  consumption: Evidence from yardstick competition theory}.
\newblock \bibinfo{journal}{Energy Econ.} \bibinfo{volume}{130},
  \bibinfo{pages}{107265}.
\newblock \DOIprefix\doi{10.1016/j.eneco.2023.107265}.
\bibitem[{Adil et~al.(2022)Adil, Bhatti, Waqar and
  Amin}]{Adil-Bhatti-Waqar-Amin-2022-JPublicAff}
\bibinfo{author}{Adil, S.}, \bibinfo{author}{Bhatti, A.A.},
  \bibinfo{author}{Waqar, S.}, \bibinfo{author}{Amin, S.},
  \bibinfo{year}{2022}.
\newblock \bibinfo{title}{Unleashing the indirect influence of oil prices on
  food prices via exchange rate: New evidence from {P}akistan}.
\newblock \bibinfo{journal}{J. Public Aff.} \bibinfo{volume}{22},
  \bibinfo{pages}{e2615}.
\newblock \DOIprefix\doi{10.1002/pa.2615}.
\bibitem[{Ahmadian-Yazdi et~al.(2024)Ahmadian-Yazdi, Roudari, Omidi, Mensi and
  Al-Yahyaee}]{AhmadianYazdi-Roudari-Omidi-Mensi-AlYahyaee-2024-IntRevEconFinanc}
\bibinfo{author}{Ahmadian-Yazdi, F.}, \bibinfo{author}{Roudari, S.},
  \bibinfo{author}{Omidi, V.}, \bibinfo{author}{Mensi, W.},
  \bibinfo{author}{Al-Yahyaee, K.H.}, \bibinfo{year}{2024}.
\newblock \bibinfo{title}{Contagion effect between fuel fossil energies and
  agricultural commodity markets and portfolio management implications}.
\newblock \bibinfo{journal}{Int. Rev. Econ. Financ.} \bibinfo{volume}{95},
  \bibinfo{pages}{103492}.
\newblock \DOIprefix\doi{10.1016/j.iref.2024.103492}.
\bibitem[{Antonakakis et~al.(2020)Antonakakis, Chatziantoniou and
  Gabauer}]{Antonakakis-Chatziantoniou-Gabauer-2020-JRiskFinancManag}
\bibinfo{author}{Antonakakis, N.}, \bibinfo{author}{Chatziantoniou, I.},
  \bibinfo{author}{Gabauer, D.}, \bibinfo{year}{2020}.
\newblock \bibinfo{title}{Refined measures of dynamic connectedness based on
  time-varying parameter vector autoregressions}.
\newblock \bibinfo{journal}{J. Risk Financ. Manag.} \bibinfo{volume}{13},
  \bibinfo{pages}{84}.
\newblock \DOIprefix\doi{10.3390/jrfm13040084}.
\bibitem[{Barun{\'{i}}k and
  K{\v{r}}hl{\'{i}}k(2018)}]{Barunik-Kehlik-2018-JFinancEconom}
\bibinfo{author}{Barun{\'{i}}k, J.}, \bibinfo{author}{K{\v{r}}hl{\'{i}}k, T.},
  \bibinfo{year}{2018}.
\newblock \bibinfo{title}{Measuring the frequency dynamics of financial
  connectedness and systemic risk}.
\newblock \bibinfo{journal}{J. Financ. Econom.} \bibinfo{volume}{16},
  \bibinfo{pages}{271--296}.
\newblock \DOIprefix\doi{10.1093/jjfinec/nby001}.
\bibitem[{Bouri et~al.(2022)Bouri, Iqbal and
  Klein}]{Bouri-Iqbal-Klein-2022-FinancResLett}
\bibinfo{author}{Bouri, E.}, \bibinfo{author}{Iqbal, N.},
  \bibinfo{author}{Klein, T.}, \bibinfo{year}{2022}.
\newblock \bibinfo{title}{Climate policy uncertainty and the price dynamics of
  green and brown energy stocks}.
\newblock \bibinfo{journal}{Financ. Res. Lett.} \bibinfo{volume}{47},
  \bibinfo{pages}{102740}.
\newblock \DOIprefix\doi{10.1016/j.frl.2022.102740}.
\bibitem[{Chen et~al.(2022)Chen, Liang, Ding and
  Liu}]{Chen-Liang-Ding-Liu-2022-EnergyEcon}
\bibinfo{author}{Chen, J.}, \bibinfo{author}{Liang, Z.}, \bibinfo{author}{Ding,
  Q.}, \bibinfo{author}{Liu, Z.}, \bibinfo{year}{2022}.
\newblock \bibinfo{title}{Extreme spillovers among fossil energy, clean energy,
  and metals markets: Evidence from a quantile-based analysis}.
\newblock \bibinfo{journal}{Energy Econ.} \bibinfo{volume}{107},
  \bibinfo{pages}{105880}.
\newblock \DOIprefix\doi{10.1016/j.eneco.2022.105880}.
\bibitem[{Diab and Karaki(2023)}]{Diab-Karaki-2023-EnergyEcon}
\bibinfo{author}{Diab, S.}, \bibinfo{author}{Karaki, M.B.},
  \bibinfo{year}{2023}.
\newblock \bibinfo{title}{Do increases in gasoline prices cause higher food
  prices?}
\newblock \bibinfo{journal}{Energy Econ.} \bibinfo{volume}{127},
  \bibinfo{pages}{107066}.
\newblock \DOIprefix\doi{10.1016/j.eneco.2023.107066}.
\bibitem[{Diebold and Yilmaz(2012)}]{Diebold-Yilmaz-2012-IntJForecast}
\bibinfo{author}{Diebold, F.X.}, \bibinfo{author}{Yilmaz, K.},
  \bibinfo{year}{2012}.
\newblock \bibinfo{title}{Better to give than to receive: {P}redictive
  directional measurement of volatility spillovers}.
\newblock \bibinfo{journal}{Int. J. Forecast.} \bibinfo{volume}{28},
  \bibinfo{pages}{57--66}.
\newblock \DOIprefix\doi{10.1016/j.ijforecast.2011.02.006}.
\bibitem[{Diebold and Yilmaz(2014)}]{Diebold-Yilmaz-2014-JEconom}
\bibinfo{author}{Diebold, F.X.}, \bibinfo{author}{Yilmaz, K.},
  \bibinfo{year}{2014}.
\newblock \bibinfo{title}{On the network topology of variance decompositions:
  {M}easuring the connectedness of financial firms}.
\newblock \bibinfo{journal}{J. Econom.} \bibinfo{volume}{182},
  \bibinfo{pages}{119--134}.
\newblock \DOIprefix\doi{10.1016/j.jeconom.2014.04.012}.
\bibitem[{Gavriilidis(2021)}]{Gavriilidis-2021-SSRN}
\bibinfo{author}{Gavriilidis, K.}, \bibinfo{year}{2021}.
\newblock \bibinfo{title}{Measuring climate policy uncertainty}.
\newblock \DOIprefix\doi{10.2139/ssrn.3847388}.
\bibitem[{Georgiou et~al.(2018)Georgiou, Acha, Shah and
  Markides}]{Georgiou-Acha-Shah-Markides-2018-JCleanProd}
\bibinfo{author}{Georgiou, S.}, \bibinfo{author}{Acha, S.},
  \bibinfo{author}{Shah, N.}, \bibinfo{author}{Markides, C.N.},
  \bibinfo{year}{2018}.
\newblock \bibinfo{title}{A generic tool for quantifying the energy
  requirements of glasshouse food production}.
\newblock \bibinfo{journal}{J. Clean Prod.} \bibinfo{volume}{191},
  \bibinfo{pages}{384--399}.
\newblock \DOIprefix\doi{10.1016/j.jclepro.2018.03.278}.
\bibitem[{Gong et~al.(2023)Gong, Jin and Liu}]{Gong-Jin-Liu-2023-Energy}
\bibinfo{author}{Gong, X.}, \bibinfo{author}{Jin, Y.}, \bibinfo{author}{Liu,
  T.}, \bibinfo{year}{2023}.
\newblock \bibinfo{title}{Analyzing pure contagion between crude oil and
  agricultural futures markets}.
\newblock \bibinfo{journal}{Energy} \bibinfo{volume}{269},
  \bibinfo{pages}{126757}.
\newblock \DOIprefix\doi{10.1016/j.energy.2023.126757}.
\bibitem[{Guo et~al.(2022)Guo, Long and
  Luo}]{Guo-Long-Luo-2022-IntRevFinancAnal}
\bibinfo{author}{Guo, J.}, \bibinfo{author}{Long, S.}, \bibinfo{author}{Luo,
  W.}, \bibinfo{year}{2022}.
\newblock \bibinfo{title}{Nonlinear effects of climate policy uncertainty and
  financial speculation on the global prices of oil and gas}.
\newblock \bibinfo{journal}{Int. Rev. Financ. Anal.} \bibinfo{volume}{83},
  \bibinfo{pages}{102286}.
\newblock \DOIprefix\doi{10.1016/j.irfa.2022.102286}.
\bibitem[{Guo et~al.(2023)Guo, Li, Zhang, Ji and
  Zhao}]{Guo-Li-Zhang-Ji-Zhao-2023-JCommodMark}
\bibinfo{author}{Guo, K.}, \bibinfo{author}{Li, Y.}, \bibinfo{author}{Zhang,
  Y.}, \bibinfo{author}{Ji, Q.}, \bibinfo{author}{Zhao, W.},
  \bibinfo{year}{2023}.
\newblock \bibinfo{title}{How are climate risk shocks connected to agricultural
  markets?}
\newblock \bibinfo{journal}{J. Commod. Mark.} \bibinfo{volume}{32},
  \bibinfo{pages}{100367}.
\newblock \DOIprefix\doi{10.1016/j.jcomm.2023.100367}.
\bibitem[{Han et~al.(2022)Han, Hua, Engel, Guan, Yin, Wu, Sun and
  Wang}]{Han-Hua-Engel-Guan-Yin-Wu-Sun-Wang-2022-AgricWaterManage}
\bibinfo{author}{Han, X.}, \bibinfo{author}{Hua, E.}, \bibinfo{author}{Engel,
  B.A.}, \bibinfo{author}{Guan, J.}, \bibinfo{author}{Yin, J.},
  \bibinfo{author}{Wu, N.}, \bibinfo{author}{Sun, S.}, \bibinfo{author}{Wang,
  Y.}, \bibinfo{year}{2022}.
\newblock \bibinfo{title}{Understanding implications of climate change and
  socio-economic development for the water-energy-food nexus: A meta-regression
  analysis}.
\newblock \bibinfo{journal}{Agric. Water Manage.} \bibinfo{volume}{269},
  \bibinfo{pages}{107693}.
\newblock \DOIprefix\doi{10.1016/j.agwat.2022.107693}.
\bibitem[{Iqbal et~al.(2024)Iqbal, Bouri, Shahzad and
  Alsagr}]{Iqbal-Bouri-Shahzad-Alsagr-2024-EnergyEcon}
\bibinfo{author}{Iqbal, N.}, \bibinfo{author}{Bouri, E.},
  \bibinfo{author}{Shahzad, S.J.H.}, \bibinfo{author}{Alsagr, N.},
  \bibinfo{year}{2024}.
\newblock \bibinfo{title}{Asymmetric impacts of {C}hinese climate policy
  uncertainty on {C}hinese asset prices}.
\newblock \bibinfo{journal}{Energy Econ.} \bibinfo{volume}{133},
  \bibinfo{pages}{107518}.
\newblock \DOIprefix\doi{10.1016/j.eneco.2024.107518}.
\bibitem[{Issariyakul and
  Dalai(2014)}]{Issariyakul-Dalai-2014-RenewSustEnergRev}
\bibinfo{author}{Issariyakul, T.}, \bibinfo{author}{Dalai, A.K.},
  \bibinfo{year}{2014}.
\newblock \bibinfo{title}{Biodiesel from vegetable oils}.
\newblock \bibinfo{journal}{Renew. Sust. Energ. Rev.} \bibinfo{volume}{31},
  \bibinfo{pages}{446--471}.
\newblock \DOIprefix\doi{10.1016/j.rser.2013.11.001}.
\bibitem[{Kang et~al.(2019)Kang, Tiwari, Albulescu and
  Yoon}]{Kang-Tiwari-Albulescu-Yoon-2019-EnergyEcon}
\bibinfo{author}{Kang, S.H.}, \bibinfo{author}{Tiwari, A.K.},
  \bibinfo{author}{Albulescu, C.T.}, \bibinfo{author}{Yoon, S.M.},
  \bibinfo{year}{2019}.
\newblock \bibinfo{title}{Exploring the time-frequency connectedness and
  network among crude oil and agriculture commodities v1}.
\newblock \bibinfo{journal}{Energy Econ.} \bibinfo{volume}{84},
  \bibinfo{pages}{104543}.
\newblock \DOIprefix\doi{10.1016/j.eneco.2019.104543}.
\bibitem[{Karim et~al.(2023)Karim, Naeem, Shafiullah, Lucey and
  Ashraf}]{Karim-Naeem-Shafiullah-Lucey-Ashraf-2023-FinancResLett}
\bibinfo{author}{Karim, S.}, \bibinfo{author}{Naeem, M.A.},
  \bibinfo{author}{Shafiullah, M.}, \bibinfo{author}{Lucey, B.M.},
  \bibinfo{author}{Ashraf, S.}, \bibinfo{year}{2023}.
\newblock \bibinfo{title}{Asymmetric relationship between climate policy
  uncertainty and energy metals: Evidence from cross-quantilogram}.
\newblock \bibinfo{journal}{Financ. Res. Lett.} \bibinfo{volume}{54},
  \bibinfo{pages}{103728}.
\newblock \DOIprefix\doi{10.1016/j.frl.2023.103728}.
\bibitem[{Kettunen et~al.(2011)Kettunen, Bunn and
  Blyth}]{Kettunen-Bunn-Blyth-2011-EnergyJ}
\bibinfo{author}{Kettunen, J.}, \bibinfo{author}{Bunn, D.W.},
  \bibinfo{author}{Blyth, W.}, \bibinfo{year}{2011}.
\newblock \bibinfo{title}{Investment propensities under carbon policy
  uncertainty}.
\newblock \bibinfo{journal}{Energy J.} \bibinfo{volume}{32},
  \bibinfo{pages}{77--117}.
\bibitem[{Khalfaoui et~al.(2022)Khalfaoui, Mefteh-Wali, Viviani, Ben~Jabeur,
  Abedin and
  Lucey}]{Khalfaoui-MeftehWali-Viviani-BenJabeur-Abedin-Lucey-2022-TechnolForecastSocChang}
\bibinfo{author}{Khalfaoui, R.}, \bibinfo{author}{Mefteh-Wali, S.},
  \bibinfo{author}{Viviani, J.L.}, \bibinfo{author}{Ben~Jabeur, S.},
  \bibinfo{author}{Abedin, M.Z.}, \bibinfo{author}{Lucey, B.M.},
  \bibinfo{year}{2022}.
\newblock \bibinfo{title}{How do climate risk and clean energy spillovers, and
  uncertainty affect {US} stock markets?}
\newblock \bibinfo{journal}{Technol. Forecast. Soc. Chang.}
  \bibinfo{volume}{185}, \bibinfo{pages}{122083}.
\newblock \DOIprefix\doi{10.1016/j.techfore.2022.122083}.
\bibitem[{Laborde et~al.(2021)Laborde, Mamun, Martin, Pineiro and
  Vos}]{Laborde-Mamun-Martin-Pineiro-Vos-2021-NatCommun}
\bibinfo{author}{Laborde, D.}, \bibinfo{author}{Mamun, A.},
  \bibinfo{author}{Martin, W.}, \bibinfo{author}{Pineiro, V.},
  \bibinfo{author}{Vos, R.}, \bibinfo{year}{2021}.
\newblock \bibinfo{title}{Agricultural subsidies and global greenhouse gas
  emissions}.
\newblock \bibinfo{journal}{Nat. Commun.} \bibinfo{volume}{12},
  \bibinfo{pages}{2601}.
\newblock \DOIprefix\doi{10.1038/s41467-021-22703-1}.
\bibitem[{Menier et~al.(2024)Menier, Bagnarosa and
  Gohin}]{Menier-Bagnarosa-Gohin-2024-ApplEcon}
\bibinfo{author}{Menier, R.}, \bibinfo{author}{Bagnarosa, G.},
  \bibinfo{author}{Gohin, A.}, \bibinfo{year}{2024}.
\newblock \bibinfo{title}{On the dependence structure of {E}uropean vegetable
  oil markets}.
\newblock \bibinfo{journal}{Appl. Econ.} \bibinfo{volume}{56},
  \bibinfo{pages}{6611--6630}.
\newblock \DOIprefix\doi{10.1080/00036846.2023.2275220}.
\bibitem[{Miljkovic and Vatsa(2023)}]{Miljkovic-Vatsa-2023-IntRevFinancAnal}
\bibinfo{author}{Miljkovic, D.}, \bibinfo{author}{Vatsa, P.},
  \bibinfo{year}{2023}.
\newblock \bibinfo{title}{On the linkages between energy and agricultural
  commodity prices: A dynamic time warping analysis}.
\newblock \bibinfo{journal}{Int. Rev. Financ. Anal.} \bibinfo{volume}{90},
  \bibinfo{pages}{102834}.
\newblock \DOIprefix\doi{10.1016/j.irfa.2023.102834}.
\bibitem[{Mohammed(2022)}]{Mohammed-2022-OPECEnergyRev}
\bibinfo{author}{Mohammed, R.}, \bibinfo{year}{2022}.
\newblock \bibinfo{title}{The impact of crude oil price on food prices in
  {I}raq}.
\newblock \bibinfo{journal}{OPEC Energy Rev.} \bibinfo{volume}{46},
  \bibinfo{pages}{106--122}.
\newblock \DOIprefix\doi{10.1111/opec.12225}.
\bibitem[{Myers et~al.(2014)Myers, Johnson, Helmar and
  Baumes}]{Myers-Johnson-Helmar-Baumes-2014-AmJAgrEcon}
\bibinfo{author}{Myers, R.J.}, \bibinfo{author}{Johnson, S.R.},
  \bibinfo{author}{Helmar, M.}, \bibinfo{author}{Baumes, H.},
  \bibinfo{year}{2014}.
\newblock \bibinfo{title}{Long-run and short-run co-movements in energy prices
  and the prices of agricultural feedstocks for biofuel}.
\newblock \bibinfo{journal}{Am. J. Agr. Econ.} \bibinfo{volume}{96},
  \bibinfo{pages}{991--1008}.
\newblock \DOIprefix\doi{10.1093/ajae/aau003}.
\bibitem[{Peri and Baldi(2010)}]{Peri-Baldi-2010-EnergyEcon}
\bibinfo{author}{Peri, M.}, \bibinfo{author}{Baldi, L.}, \bibinfo{year}{2010}.
\newblock \bibinfo{title}{Vegetable oil market and biofuel policy: {A}n
  asymmetric cointegration approach}.
\newblock \bibinfo{journal}{Energy Econ.} \bibinfo{volume}{32},
  \bibinfo{pages}{687--693}.
\newblock \DOIprefix\doi{10.1016/j.eneco.2009.09.004}.
\bibitem[{Qiao et~al.(2024)Qiao, Chang, Mai and
  Dang}]{Qiao-Chang-Mai-Dang-2024-EnergyEcon}
\bibinfo{author}{Qiao, S.}, \bibinfo{author}{Chang, Y.}, \bibinfo{author}{Mai,
  X.X.}, \bibinfo{author}{Dang, Y.J.}, \bibinfo{year}{2024}.
\newblock \bibinfo{title}{Climate policy uncertainty, clean energy and energy
  metals: {A} quantile time-frequency spillover study}.
\newblock \bibinfo{journal}{Energy Econ.} \bibinfo{volume}{139},
  \bibinfo{pages}{107919}.
\newblock \DOIprefix\doi{10.1016/j.eneco.2024.107919}.
\bibitem[{Saeed et~al.(2021)Saeed, Bouri and
  Alsulami}]{Saeed-Bouri-Alsulami-2021-EnergyEcon}
\bibinfo{author}{Saeed, T.}, \bibinfo{author}{Bouri, E.},
  \bibinfo{author}{Alsulami, H.}, \bibinfo{year}{2021}.
\newblock \bibinfo{title}{Extreme return connectedness and its determinants
  between clean/green and dirty energy investments}.
\newblock \bibinfo{journal}{Energy Econ.} \bibinfo{volume}{96},
  \bibinfo{pages}{105017}.
\newblock \DOIprefix\doi{10.1016/j.eneco.2020.105017}.
\bibitem[{Shang et~al.(2022)Shang, Han, Gozgor, Mahalik and
  Sahoo}]{Shang-Han-Gozgor-Mahalik-Sahoo-2022-RenewEnergy}
\bibinfo{author}{Shang, Y.}, \bibinfo{author}{Han, D.},
  \bibinfo{author}{Gozgor, G.}, \bibinfo{author}{Mahalik, M.K.},
  \bibinfo{author}{Sahoo, B.K.}, \bibinfo{year}{2022}.
\newblock \bibinfo{title}{The impact of climate policy uncertainty on renewable
  and non-renewable energy demand in the united states}.
\newblock \bibinfo{journal}{Renew. Energy} \bibinfo{volume}{197},
  \bibinfo{pages}{654--667}.
\newblock \DOIprefix\doi{10.1016/j.renene.2022.07.159}.
\bibitem[{Siddique et~al.(2023)Siddique, Nobanee, Hasan, Uddin, Hossain and
  Park}]{Siddique-Nobanee-Hasan-Uddin-Hossain-Park-2023-EnergyEcon}
\bibinfo{author}{Siddique, M.A.}, \bibinfo{author}{Nobanee, H.},
  \bibinfo{author}{Hasan, M.B.}, \bibinfo{author}{Uddin, G.S.},
  \bibinfo{author}{Hossain, M.N.}, \bibinfo{author}{Park, D.},
  \bibinfo{year}{2023}.
\newblock \bibinfo{title}{How do energy markets react to climate policy
  uncertainty? {F}ossil vs. renewable and low-carbon energy assets}.
\newblock \bibinfo{journal}{Energy Econ.} \bibinfo{volume}{128},
  \bibinfo{pages}{107195}.
\newblock \DOIprefix\doi{10.1016/j.eneco.2023.107195}.
\bibitem[{Su et~al.(2022)Su, Yuan, Tao and
  Shao}]{Su-Yuan-Tao-Shao-2022-TechnolForecastSocChang}
\bibinfo{author}{Su, C.W.}, \bibinfo{author}{Yuan, X.}, \bibinfo{author}{Tao,
  R.}, \bibinfo{author}{Shao, X.}, \bibinfo{year}{2022}.
\newblock \bibinfo{title}{Time and frequency domain connectedness analysis of
  the energy transformation under climate policy}.
\newblock \bibinfo{journal}{Technol. Forecast. Soc. Chang.}
  \bibinfo{volume}{184}, \bibinfo{pages}{121978}.
\newblock \DOIprefix\doi{10.1016/j.techfore.2022.121978}.
\bibitem[{Syed et~al.(2023)Syed, Apergis and
  Goh}]{Syed-Apergis-Goh-2023-Energy}
\bibinfo{author}{Syed, Q.R.}, \bibinfo{author}{Apergis, N.},
  \bibinfo{author}{Goh, S.K.}, \bibinfo{year}{2023}.
\newblock \bibinfo{title}{The dynamic relationship between climate policy
  uncertainty and renewable energy in the {US}: Applying the novel fourier
  augmented autoregressive distributed lags approach}.
\newblock \bibinfo{journal}{Energy} \bibinfo{volume}{275},
  \bibinfo{pages}{127383}.
\newblock \DOIprefix\doi{10.1016/j.energy.2023.127383}.
\bibitem[{Taghizadeh-Hesary et~al.(2019)Taghizadeh-Hesary, Rasoulinezhad and
  Yoshino}]{TaghizadehHesary-Rasoulinezhad-Yoshino-2019-EnergyPolicy}
\bibinfo{author}{Taghizadeh-Hesary, F.}, \bibinfo{author}{Rasoulinezhad, E.},
  \bibinfo{author}{Yoshino, N.}, \bibinfo{year}{2019}.
\newblock \bibinfo{title}{Energy and food security: Linkages through price
  volatility}.
\newblock \bibinfo{journal}{Energy Policy} \bibinfo{volume}{128},
  \bibinfo{pages}{796--806}.
\newblock \DOIprefix\doi{10.1016/j.enpol.2018.12.043}.
\bibitem[{Tanaka et~al.(2023)Tanaka, Guo and
  Wang}]{Tanaka-Guo-Wang-2023-GCBBioenergy}
\bibinfo{author}{Tanaka, T.}, \bibinfo{author}{Guo, J.}, \bibinfo{author}{Wang,
  X.}, \bibinfo{year}{2023}.
\newblock \bibinfo{title}{Price interconnection of fuel and food markets:
  Evidence from biodiesel in the {U}nited {S}tates}.
\newblock \bibinfo{journal}{GCB Bioenergy} \bibinfo{volume}{15},
  \bibinfo{pages}{886--899}.
\newblock \DOIprefix\doi{10.1111/gcbb.13055}.
\bibitem[{Tanaka et~al.(2024)Tanaka, Guo and
  Wang}]{Tanaka-Guo-Wang-2024-EnergyEcon}
\bibinfo{author}{Tanaka, T.}, \bibinfo{author}{Guo, J.}, \bibinfo{author}{Wang,
  X.}, \bibinfo{year}{2024}.
\newblock \bibinfo{title}{Understanding the spillover effects of ethanol
  production and energy prices on {A}frican food markets: A time-varying
  approach}.
\newblock \bibinfo{journal}{Energy Econ.} \bibinfo{volume}{134},
  \bibinfo{pages}{107583}.
\newblock \DOIprefix\doi{10.1016/j.eneco.2024.107583}.
\bibitem[{Vatsa et~al.(2023)Vatsa, Miljkovic and
  Baek}]{Vatsa-Miljkovic-Baek-2023-JAgricEcon}
\bibinfo{author}{Vatsa, P.}, \bibinfo{author}{Miljkovic, D.},
  \bibinfo{author}{Baek, J.}, \bibinfo{year}{2023}.
\newblock \bibinfo{title}{Linkages between natural gas, fertiliser and cereal
  prices: A note}.
\newblock \bibinfo{journal}{J. Agric. Econ.} \bibinfo{volume}{74},
  \bibinfo{pages}{935--940}.
\newblock \DOIprefix\doi{10.1111/1477-9552.12532}.
\bibitem[{Vo and Tran(2024)}]{Vo-Tran-2024-TechnolForecastSocChang}
\bibinfo{author}{Vo, D.H.}, \bibinfo{author}{Tran, M.P.B.},
  \bibinfo{year}{2024}.
\newblock \bibinfo{title}{Volatility spillovers between energy and agriculture
  markets during the ongoing food \& energy crisis: Does uncertainty from the
  {R}usso-{U}krainian conflict matter?}
\newblock \bibinfo{journal}{Technol. Forecast. Soc. Chang.}
  \bibinfo{volume}{208}, \bibinfo{pages}{123723}.
\newblock \DOIprefix\doi{10.1016/j.techfore.2024.123723}.
\bibitem[{Wang et~al.(2023)Wang, Wang, Yunis and
  Kchouri}]{Wang-Wang-Yunis-Kchouri-2023-EnergyEcon}
\bibinfo{author}{Wang, K.H.}, \bibinfo{author}{Wang, Z.S.},
  \bibinfo{author}{Yunis, M.}, \bibinfo{author}{Kchouri, B.},
  \bibinfo{year}{2023}.
\newblock \bibinfo{title}{Spillovers and connectedness among climate policy
  uncertainty, energy, green bond and carbon markets: A global perspective}.
\newblock \bibinfo{journal}{Energy Econ.} \bibinfo{volume}{128},
  \bibinfo{pages}{107170}.
\newblock \DOIprefix\doi{10.1016/j.eneco.2023.107170}.
\bibitem[{Wang(2020)}]{Wang-2020-EnergyEcon}
\bibinfo{author}{Wang, X.}, \bibinfo{year}{2020}.
\newblock \bibinfo{title}{Frequency dynamics of volatility spillovers among
  crude oil and international stock markets: The role of the interest rate}.
\newblock \bibinfo{journal}{Energy Econ.} \bibinfo{volume}{91},
  \bibinfo{pages}{104900}.
\newblock \DOIprefix\doi{10.1016/j.eneco.2020.104900}.
\bibitem[{Wu et~al.(2023)Wu, Ren, Wan and
  Liu}]{Wu-Ren-Wan-Liu-2023-FinancResLett}
\bibinfo{author}{Wu, Y.}, \bibinfo{author}{Ren, W.}, \bibinfo{author}{Wan, J.},
  \bibinfo{author}{Liu, X.}, \bibinfo{year}{2023}.
\newblock \bibinfo{title}{Time-frequency volatility connectedness between
  fossil energy and agricultural commodities: Comparing the {COVID}-19 pandemic
  with the {R}ussia-{U}kraine conflict}.
\newblock \bibinfo{journal}{Financ. Res. Lett.} \bibinfo{volume}{55},
  \bibinfo{pages}{103866}.
\newblock \DOIprefix\doi{10.1016/j.frl.2023.103866}.
\bibitem[{Yan and Cheung(2023)}]{Yan-Cheung-2023-FinancResLett}
\bibinfo{author}{Yan, W.L.}, \bibinfo{author}{Cheung, A.W.K.},
  \bibinfo{year}{2023}.
\newblock \bibinfo{title}{The dynamic spillover effects of climate policy
  uncertainty and coal price on carbon price: Evidence from {C}hina}.
\newblock \bibinfo{journal}{Financ. Res. Lett.} \bibinfo{volume}{53},
  \bibinfo{pages}{103400}.
\newblock \DOIprefix\doi{10.1016/j.frl.2022.103400}.
\bibitem[{Yoon(2022)}]{Yoon-2022-RenewEnergy}
\bibinfo{author}{Yoon, S.M.}, \bibinfo{year}{2022}.
\newblock \bibinfo{title}{On the interdependence between biofuel, fossil fuel
  and agricultural food prices: Evidence from quantile tests}.
\newblock \bibinfo{journal}{Renew. Energy} \bibinfo{volume}{199},
  \bibinfo{pages}{536--545}.
\newblock \DOIprefix\doi{10.1016/j.renene.2022.08.136}.
\bibitem[{Youssef and Mokni(2021)}]{Youssef-Mokni-2021-IntEconJ}
\bibinfo{author}{Youssef, M.}, \bibinfo{author}{Mokni, K.},
  \bibinfo{year}{2021}.
\newblock \bibinfo{title}{On the nonlinear impact of oil price shocks on the
  world food prices under different markets conditions}.
\newblock \bibinfo{journal}{Int. Econ. J.} \bibinfo{volume}{35},
  \bibinfo{pages}{73--95}.
\newblock \DOIprefix\doi{10.1080/10168737.2020.1870524}.
\bibitem[{Yu et~al.(2023)Yu, Peng, Zakaria, Mahmood and
  Khalid}]{Yu-Peng-Zakaria-Mahmood-Khalid-2023-JBusEconManag}
\bibinfo{author}{Yu, Y.}, \bibinfo{author}{Peng, C.}, \bibinfo{author}{Zakaria,
  M.}, \bibinfo{author}{Mahmood, H.}, \bibinfo{author}{Khalid, S.},
  \bibinfo{year}{2023}.
\newblock \bibinfo{title}{Nonlinear effects of crude oil dependency on food
  prices in {C}hina: Evidence from quantile-on-quantile approach}.
\newblock \bibinfo{journal}{J. Bus. Econ. Manag.} \bibinfo{volume}{24},
  \bibinfo{pages}{696--711}.
\newblock \DOIprefix\doi{10.3846/jbem.2023.20192}.
\bibitem[{Zhou et~al.(2023)Zhou, Siddik, Guo and
  Li}]{Zhou-Siddik-Guo-Li-2023-RenewEnergy}
\bibinfo{author}{Zhou, D.}, \bibinfo{author}{Siddik, A.B.},
  \bibinfo{author}{Guo, L.}, \bibinfo{author}{Li, H.}, \bibinfo{year}{2023}.
\newblock \bibinfo{title}{Dynamic relationship among climate policy
  uncertainty, oil price and renewable energy consumption-findings from
  {TVP-SV-VAR} approach}.
\newblock \bibinfo{journal}{Renew. Energy} \bibinfo{volume}{204},
  \bibinfo{pages}{722--732}.
\newblock \DOIprefix\doi{10.1016/j.renene.2023.01.018}.

\end{thebibliography}

\end{document}